\documentclass[12pt]{elsarticle}

\usepackage{hyperref}

\usepackage{amssymb,amsmath}
\usepackage{bm}
\usepackage{epstopdf}

\allowdisplaybreaks

\begin{document}
    \begin{frontmatter}
        \title{Photoproduction of $e^+e^-$ pair in a Coulomb field near the threshold.}
        \author{P.A. Krachkov}
        \author{R.N. Lee}
        \author{A.I. Milstein}
        
        \address{Budker Institute of Nuclear Physics, SB RAS, Novosibirsk, 630090, Russia}
        
        \begin{abstract}
            We consider the electron-positron pair production by a photon in Coulomb field near threshold. We obtain the analytical results for the particle energy spectrum and angular distribution exactly in the parameter $Ze^2/\hbar v$.
            We show that Furry-Sommerfeld-Maue approximation for the wave functions used by Nishina, Tomonaga and Sakata in Ref. \cite{NTS1934} is not sufficient and leads to wrong results.
        \end{abstract}
    \end{frontmatter}
    
%
%
%
%
%
%
%
%
%
%
%
%

\section{Introduction}
Bremsstrahlung and photoproduction in an atomic field are among the basic processes of quantum electrodynamics.  Their first studies were performed in just a few years after the discovery of quantum mechanics. Not only their Born cross sections but also the nonrelativistic limits exact in the parameter $Ze^2/\hbar v$ have been considered already in early 1930s. In particular, the spectrum of the nonrelativistic electron bremsstrahlung in a Coulomb field has been calculated in 1931 by Sommerfeld \cite{Sommerfeld1931} exactly in the parameters $\eta_{i,f}=Z\alpha/v_{i,f}$ (here $Z$ is the atomic charge number, $\alpha=e^2$ is the fine-structure constant, $e$ is the electron charge, $v_{i,f}\ll 1$ are  velocities of initial and final electrons, $ Z\alpha\ll 1$, $\hbar=c=1$). 
The derivation of Ref. \cite{Sommerfeld1931} (see also  \cite[\S92]{BLP1982}) is based on the fact that the main contribution to the matrix element comes from the distances that are much greater than the Compton wavelength $\lambda_C=1/m$ ($m$ is the electron mass), so that the relativistic corrections may be neglected. Note that the resulting expression is a rather non-trivial function of $\eta_{i,f}$. 

A completely different situation arises when calculating the cross section of $e^+e^-$ pair photoproduction in a Coulomb field in the near-threshold region,   $\omega-2m\ll m$ ($\omega$ is the photon energy). 
Although in this case the electron and positron velocities $v_p$ and $v_q$ are small compared to the speed of light, the characteristic distances that make the main contribution to the photoproduction amplitude are of the order of $\lambda_C$. 
Therefore, the nonrelativistic approximation for the wave functions cannot be used.
Unfortunately, the exact solution of the Dirac equation for a relativistic particle in a Coulomb field can only be written as a sum over angular momenta $j$ which can not be taken in closed form, in contrast to the case of the Schr\"odinger equation in a Coulomb field. 
However, the sum can be taken if one makes a substitution $\gamma_L=\sqrt{L^2-(Z\alpha)^2}\to L=j+1/2=1,2,\ldots$, which corresponds to neglecting $Z\alpha$ as compared to $L$. 
The resulting expression is the celebrated Furry-Sommerfeld-Maue (FSM) wave function  \cite{Furry1934,Sommerfeld1935}. It is this approximation which was used in Ref. \cite{NTS1934} to obtain the cross section of $e^+e^-$ pair photoproduction in a Coulomb field in two limiting cases: for high and for low energies of electron and positron. 
It was assumed that the parameter $Z\alpha$ is small compared to unity which justifies the applicability of FSM approximation. The applicability was supported also by the agreement of the obtained high-energy spectrum with the result of Bethe and Heitler  \cite{Bethe1934}. 
Note that the scattering angles $\theta_\pm$ of the produced particles are small in the high-energy limit. Then the characteristic angular momenta are large,  $L\sim 1/\theta_\pm\gg 1$, and the requirement $Z\alpha\ll 1$ may be relaxed. This fact was later used in Ref. \cite{Davies1954} to calculate the high-energy pair photoproduction cross section exactly in $Z\alpha$. 
As to the near-threshold region (low energies of $e^\pm$), the results of Ref. \cite{NTS1934} for the angular distribution and spectrum had the correct asymptotics at $Z\alpha\ll v_{p,q}$ being in agreement with the low-energy limit of the Born cross section.

Later on, remarkable efforts have been undertaken in Refs. \cite{Overbo1968,Overbo1973} to calculate the pair photoproduction cross section exactly in both $Z\alpha$ and energies using the angular momentum sum representation of the wave function. The obtained analytic expression has a form of multiple sums with the summand expressed via Appell $F_2$ double hypergeometric function. Due to this complicated form, the numerical results obtained in Refs. \cite{Overbo1968,Overbo1973} has been restricted by photon energy  $\lesssim5$MeV. A much later and only modestly successful attempt to extend the numerical results to higher energies was undertaken in Ref. \cite{SudSharma2006}. Besides, this complicated form was probably the reason why the authors of Ref \cite{Overbo1973} did not manage to obtain the threshold asymptotics of their expressions analytically. Instead, they compared their result numerically with analytic formula of Ref.~ \cite{NTS1934}. The numerical disagreement found was ascribed to the factor $2\pi$ allegedly overlooked in Ref. \cite{NTS1934}. However, we show in the present paper that the reason of the disagreement is not an arithmetic error or typo. It turns out that the FSM approximation is not sufficient for the calculation of $e^+e^-$ pair photoproduction  cross section in a Coulomb field in the near-threshold region. Consequently,  the near-threshold results of Ref. \cite{NTS1934} which follow from FSM approximation are not correct.  The reason is that, when calculating the matrix element of the process in the near-threshold region, there is a cancellation of the leading terms of  expansion in the parameters $v_{p,q}\ll 1$. Therefore, for $Z\alpha\sim v_{p,q}$ it is necessary to take into account the first correction of the expansion of $\gamma_L$ in $(Z\alpha)^2$, i.e., to make  the replacement $\gamma_L\approx L-(Z\alpha )^2/(2L)$. We show that the account for this correction changes both the shape of angular distribution and the spectrum of the produced particles. We therefore obtain for the first time the correct analytical results for the pair photoproduction spectrum and angular distribution valid in the region $Z\alpha\sim v_{p,q}\ll 1$.

\section{Matrix element of the process in the near-threshold region}
The differential cross section of $e^+e^-$ photoproduction   has the form \cite{BLP1982}
\begin{align}\label{sec0}
	&d\sigma= \dfrac{\alpha pq\varepsilon_p \varepsilon_q\,d\varepsilon_p \,d\Omega_{\bm p}\,d\Omega_{\bm q} }{\omega(2\pi)^4}\,|T|^2\,,\nonumber\\
	&T=\int\,d^3 r\,e^{i\bm k\cdot\bm r} \,\bar U_{\bm p }^{(-)}(\bm r )\,\bm\gamma\cdot\bm e\,V_{{\bm q}}^{(+)}({\bm r})\,,
\end{align}
where $\bm p$ and $\bm q$ are the electron and positron momenta, $\varepsilon_p $ and $\varepsilon_q=\omega-\varepsilon_p$ are their energies, $\bm e$ is the photon polarization, $\bm k$ is the photon momentum, $V_{{\bm p}}^{(+)}({\bm r})$ is the negative-frequency solution of the Dirac equation in the Coulomb field, which at large distances contains a plane wave and a diverging spherical wave, while the positive-frequency solution $U_{{\bm p}}^{(-)}({\bm r})$  at large distances contains a plane wave and a converging spherical wave.
 
In order to calculate the matrix element it is convenient to express the wave functions via the large-distance asymptotics of the Green's functions of the Dirac equation, see, e.g., Ref. \cite{LMS2012}. We have
\begin{align}\label{UV}
&\overline{U}_{\bm p}^{(-)}(\bm r )= -\lim_{ r_2\to \infty }\frac{2\pi r_2}{\varepsilon_p}  e^{-ipr_2-i\eta_p\ln(2pr_2)}\bar u_{\bm p }\gamma^0 G({\bm n}_p r_2,{\bm r}|\varepsilon_{ p}) \nonumber\\
&\mspace{210mu}=\sqrt{\dfrac{\varepsilon_p+m}{2\varepsilon_p}}\left(\varphi^{+}\,f_1,\,-\varphi^{+}\dfrac{\bm\sigma\cdot\bm p}{\varepsilon_p+m}f_2\right)\,,    
\nonumber\\
&V_{{\bm q}}^{(+)}({\bm r})=\lim_{ r_1\to \infty }\frac{2\pi r_1}{\varepsilon_q}  e^{-iqr_1+i\eta_q\ln(2qr_1)}G({\bm r},{\bm n}_q r_1|-\varepsilon_{ q})\gamma^0v_{{\bm q}}\nonumber\\
&\mspace{290mu}=\sqrt{\dfrac{\varepsilon_q+m}{2\varepsilon_q}}
\begin{pmatrix}
	\tilde{f}_1\,\dfrac{{\bm \sigma}\cdot {\bm q}}{\varepsilon_q+m}\chi\\
\tilde{f}_2	\chi
\end{pmatrix}\, ,
\nonumber\\
&f_{1,2}=\sum_{L=1}^\infty\left\{\left[R_{-}A_L(x_p)\pm i\dfrac{m\eta_p}{\varepsilon_p}R_{+}B_L(x_p)\right]M_1-iR_{+}B_L(x_p)M_2\right\}\,,\, \nonumber\\
&\tilde{f}_{1,2}=\sum_{L'=1}^\infty\left\{\left[\widetilde{R}_-A_{L'}(x_q)
\pm i\dfrac{m\eta_q}{\varepsilon_q}\widetilde{R}_+B_{L'}(x_q)\right]\widetilde{M}_1-i\widetilde{R}_+B_{L'}(x_q)\widetilde{M}_2\right\},
\end{align}
where $\varphi$ and $\chi$ are two-component spinors,  $v_{{\bm q}}$ and $u_{\bm p}$ are free Dirac bispinors, and the following notation is introduced
\begin{align}
& x_{p,q} =	\bm n\cdot\bm n_{p,q}\,,\quad R_{\pm}=1\pm (\bm\sigma\cdot\bm n_{\bm p})(\bm\sigma\cdot\bm n)\,,\quad \widetilde{R}_{\pm}=1\pm (\bm\sigma\cdot\bm n)(\bm\sigma\cdot\bm n_{\bm q})\,,\nonumber\\
&A_L(x)=L\dfrac{d}{dx}[P_L(x)+P_{L-1}(x)]\,,\quad B_L(x)=\dfrac{d}{dx}[P_L(x)-P_{L-1}(x)]\,,\nonumber\\
&M_1=\dfrac{(2pr)^{\gamma_L-1}\Gamma(\gamma_L-i\eta_p)}{\Gamma(2\gamma_L+1)}\exp\left[ipr+i\frac{\pi}{2}(1-\gamma_L-i\eta_p)\right]\nonumber\\
&\mspace{210mu}\times F(\gamma_L-i\eta_p,\,2\gamma_L+1, -2ipr )\,,\nonumber\\
&M_2=\dfrac{(2pr)^{\gamma_L-1}\Gamma(\gamma_L-i\eta_p+1)}{\Gamma(2\gamma_L+1)}\exp\left[ipr+i\frac{\pi}{2}(2-\gamma_L-i\eta_p)\right]\nonumber\\
&\mspace{210mu}\times F(\gamma_L-i\eta_p+1,\,2\gamma_L+1, -2ipr )\,.
\end{align}
Here $\bm n_{\bm p}=\bm p/p$, $\bm n_{\bm q}=\bm q/p$, $\bm n=\bm r/r$, $F( a,b,x)$ is the confluent hypergeometric function, $\Gamma(x)$ is Euler gamma function, $\gamma_L=\sqrt{L^2-(Z\alpha)^2}$, $\eta_p =Z\alpha\varepsilon_p/p$, $\eta_q=Z\alpha\varepsilon_q/q$. The functions $\widetilde{M}_{1,2}$ are obtained from ${M}_{1,2}$ by replacing $p\to q,\,L\to L',\, \eta_p\to-\eta_q$.

\begin{figure}[t]
	\includegraphics[width=350\unitlength]{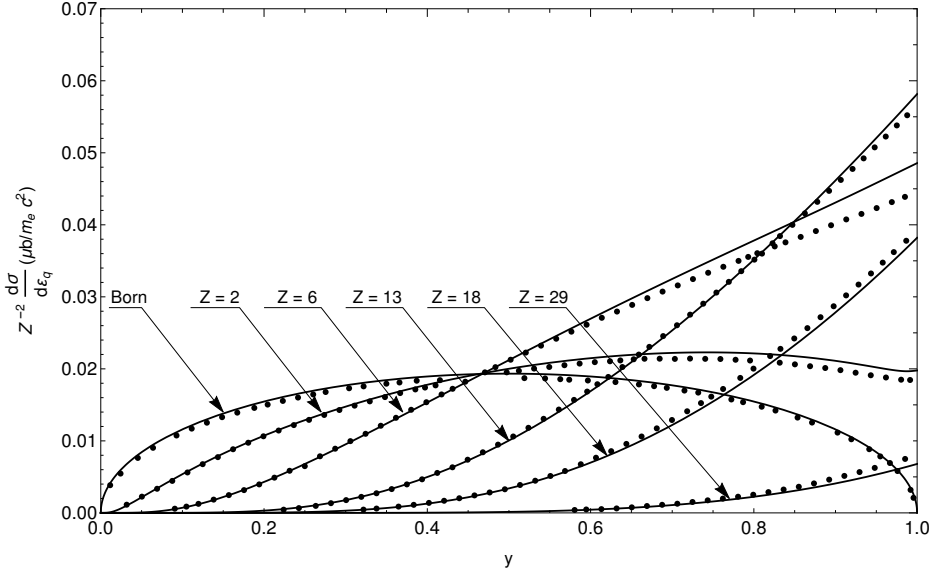}
	\caption{Dependence of $Z^{-2}\dfrac{d\sigma}{d\varepsilon_q}$ on $y=\dfrac{\varepsilon_q-m}{\omega-2m}$ for  different  $Z$ at $\omega=2.01 m$. Solid curve corresponds to our result \eqref{eq:result}, points are extracted from the plot on Fig. $1$ in Ref. \cite{Overbo1973}.}\label{pic}
\end{figure}

In the near-threshold region, the main contribution to the matrix element is determined by the distances $r\sim 1/\omega$. Therefore, we expand the wave functions in $pr,qr\ll 1$ and then make the replacements 
\begin{equation}
    \gamma_L\to L-\underline{(Z\alpha)^ 2/2L},\qquad
    \gamma_L'\to L'-\underline{(Z\alpha)^ 2/2L'}.
\end{equation}
We need to take into account the underlined  corrections in the expansion of $\gamma_{L,L'}$ because of cancellation of the leading contributions to the matrix element. Obviously, due to the factors $(pr)^{\gamma_L}$ and $(qr)^{\gamma_{L'}}$, the main contribution to the sum over $L$ and $L'$ is given only by the first few values of $ L$ and $L'$. It is easy to see that these are $(L,\,L')=(1,1),\,(1,2),\, (2,1)$.
It is also convenient to use the circular polarization of the photon. As a result of a rather cumbersome calculation, we obtain in the near-threshold region
\begin{multline}\label{eq:amplitude}
    T=\phi^+{\hat T}\chi\,,\quad {\hat T}=\dfrac{\pi\,v_pv_q}{2}\exp\left[\frac{\pi}{2}(\eta_p-\eta_q)\right]\,\Gamma(1-i\eta_p)\Gamma(1+i\eta_q)\\
    \times\bigg\{\frac{1}{4}\pi\eta_p\eta_q(\bm\sigma\cdot\bm e_\lambda)
    +\eta_p\,(\bm\sigma\cdot\bm e_\lambda)[(1+i\eta_q)(\bm n_{\bm q}\cdot\bm n_{\bm k})+\lambda\,(1+i\eta_q/2)(\bm\sigma\cdot\bm n_{\bm q})]\\
    -\eta_q[(1-i\eta_p)(\bm n_{\bm p}\cdot\bm n_{\bm k})-\lambda\,(1-i\eta_p/2)(\bm\sigma\cdot\bm n_{\bm p})](\bm\sigma\cdot\bm e_\lambda)\\
    +\underline{
    \frac{1}2 \eta_p\eta_q\left[\pi(\bm\sigma\cdot\bm e_\lambda)+i\lambda(\bm\sigma\cdot\bm e_\lambda)(\bm\sigma\cdot\bm n_{\bm q})-i\lambda(\bm\sigma\cdot\bm n_{\bm p})(\bm\sigma\cdot\bm e_\lambda)\right]    
    }\bigg\}\,.
\end{multline}
Here $\bm n_{\bm k}=\bm k/k$ and $\bm e_\lambda$ is the photon polarization vector with helicity $\lambda=\pm 1$. The underlined terms come from the expansions of $\gamma_{L,L'}$. 
After averaging over photon helicities and summing over the polarization of  electron and positron, we find the differential cross section in the near-threshold region
\begin{multline}\label{eq:resultAD}
	d\sigma=\dfrac{\alpha(Z\alpha)^2v_{\bm p}v_{\bm q}\,d\varepsilon_p \,d\Omega_{\bm p}\,d\Omega_{\bm q}}{16(2\pi)^2m^3}\dfrac{(2\pi\eta_p)(2\pi\eta_q)}{(1-e^{-2\pi\eta_p})(e^{2\pi\eta_q}-1)}\\
	\times\Big\{(1+\eta_p^2)[\bm v_{\bm p}\times\bm n_{\bm k}]^2+(1+\eta_q^2)[\bm v_{\bm q}\times\bm n_{\bm k}]^2+\dfrac{9\pi^2}{16}v_{\bm p}v_{\bm q}\eta_p\eta_q\Big\}\,.
\end{multline}
Integrating over the angles of electron and positron momenta, we find the spectrum of produced particles
\begin{equation}\label{eq:result}
	d\sigma=\dfrac{\alpha(Z\alpha)^2v_{\bm p}v_{\bm q}\,d\varepsilon_p }{6m^3}\dfrac{(2\pi\eta_p)(2\pi\eta_q)}{(1-e^{-2\pi\eta_p})(e^{2\pi\eta_q}-1)}
	\left[v_{\bm p}^2+v_{\bm q}^2+\left(\dfrac{27\pi^2}{32}+2\right)(Z\alpha)^2\right]\,.
\end{equation}
If we omit the underlined terms in Eq. \eqref{eq:amplitude} then we reproduce the (incorrect) results of Nishina, Tomonaga and Sakata, Ref. \cite{NTS1934}, for the differential cross section,
\begin{multline}\label{eq:NTSresultAD}
	d\sigma_{\text{NTS}}=\dfrac{\alpha(Z\alpha)^2v_{\bm p}v_{\bm q}\,d\varepsilon_p \,d\Omega_{\bm p}\,d\Omega_{\bm q}}{16(2\pi)^2m^3}\dfrac{(2\pi\eta_p)(2\pi\eta_q)}{(1-e^{-2\pi\eta_p})(e^{2\pi\eta_q}-1)}\\
	\times\Big\{[\bm v_{\bm p}\times\bm n_{\bm k}]^2+[\bm v_{\bm q}\times\bm n_{\bm k}]^2
	+\dfrac{1}{2}\Big[\left(\dfrac{\pi^2}{8}+1\right)v_{\bm p}v_{\bm q}+(\bm v_{\bm p}\cdot\bm n_{\bm k})(\bm v_{\bm q}\cdot\bm n_{\bm k})\Big]\eta_p\eta_q\,\Big\}\,
\end{multline}
and for the spectrum,
\begin{equation}\label{eq:NTSresult}
	d\sigma_{\text{NTS}}=\frac{\alpha(Z\alpha)^2v_{\bm p}v_{\bm q}\,d\varepsilon_p }{6m^3}\dfrac{(2\pi\eta_p)(2\pi\eta_q)}{(1-e^{-2\pi\eta_p})(e^{2\pi\eta_q}-1)}
	\left[v_{\bm p}^2+v_{\bm q}^2+\,\left(\dfrac{3\pi^2}{32}+\dfrac{3}{4}\right)\,(Z\alpha)^2\right]\,.
\end{equation}
It is remarkable that the ratio of the coefficients in front of $(Z\alpha)^2$ in square brackets in Eqs. \eqref{eq:result} and \eqref{eq:NTSresult} is $\left.(\frac{27 \pi ^2}{32}+2)\middle/(\frac{3 \pi ^2}{32}+\frac{3}{4})\right.\approx6.16$ which is indeed quite close to a factor $2\pi$ used in Ref. \cite{Overbo1968} to explain the discrepancy with Ref. \cite{NTS1934}. Note that our analytical expression \eqref{eq:result}  for the spectrum agrees well with the numerical results of \cite{Overbo1968,Overbo1973} at small energies, see Fig.\ref{pic}.

%

\section*{Conclusion}

We have considered the electron-positron photoproduction in a Coulomb field in the near-threshold region. Particle spectrum and angular distributions are obtained exactly in the parameters $\eta_{p,q}=Z\alpha/ v_{p,q}$ in an analytic form, Eqs. \eqref{eq:result} and \eqref{eq:resultAD}.  Previous results for the spectrum and angular distributions  obtained with the use of the Furry-Sommerfeld-Maue approximation, are shown to be incorrect due to the necessity to take also into account the first correction to this approximation. Our analytical results agree well with the numerical estimates. 


\bibliographystyle{apsrev4-1}
\bibliography{klm2022}

\end{document}